\documentclass[aps,prb,twocolumn,showpacs]{revtex4-1}
\usepackage{epsfig}
\usepackage{amsmath}
\usepackage{amssymb}
\usepackage{colortbl}

\begin{document}

\title{On the Area Law for \\
Disordered Free Fermions}
\author{L.~Pastur}
\author{V.~Slavin}
\affiliation{B.I. Verkin Institute for Low Temperatures and
Engineering, 61103, Kharkiv, Ukraine}
\date{\today }

\begin{abstract}
We study theoretically and numerically the entanglement entropy of
the $d$-dimensional free fermions whose one body Hamiltonian is
the Anderson model. Using basic facts of the exponential Anderson
localization, we show first that the disorder averaged entanglement entropy $%
\langle S_\Lambda \rangle$ of the $d$ dimension cube $\Lambda$ of
side length $l$ admits the area law scaling $\langle S_\Lambda
\rangle \sim l^{(d-1)}, \ l  \gg 1$ even in the gapless case,
thereby manifesting the area law in the mean for our model.
%This has to be compared with results of Ref. \onlinecite{Re-Mo:09},
%according to which the area
%law for the averaged entropy is violated at critical
%points of disordered spin chains.
For $d=1$ and $l\gg 1$ we obtain then asymptotic bounds for the
entanglement entropy of typical realizations of disorder and use
them to show that the entanglement entropy is not selfaveraging,
i.e., has non vanishing random fluctuations even if $l \gg 1$.
\end{abstract}


\maketitle

%\section{Introduction}

Entanglement is a basic ingredient of quantum description having a
great potential for applications \cite{Ho-Co:09}. An important
quantifier of entanglement is the von Neumann entropy. In the
bipartite setting, where the system is the union of a subsystem
and its environment of the characteristic lengths $l $ and $L$,
the entropy of the reduced density matrix of the subsystem
(entanglement or block entropy) may have an unusual asymptotic
behavior as a function of $l,\;1\ll l\ll L,$ if the whole system
is in its ground state. Namely, it was shown in a number of works
that the entanglement entropy is proportional to the surface area
$l^{d-1}$ of the subsystem but not to its volume $l^{d}$. The
latter (extensive) length scaling is standard in quantum
statistical mechanics for non-zero temperature (thermal entanglement,
while the former was found first in cosmology and then in other fields and is known
as the \textit{area law}. Moreover, the area law is not always valid, e.g.,
at quantum critical points of several one-dimensional (1d) translation invariant quantum
spin chains, where the entropy is proportional to $\log l,\ l\gg
1$.
%(recall that the area law in the one-dimensional case is just the
%boundedness %of the entanglement entropy in $l)$.
It is also believed and found for simple translation invariant models that a
multidimensional analog of the above divergence is $l^{d-1}\log l$ \cite%
{Am-Co:08,Gi-Kl:06}.

More generally, the area law scaling $l^{d-1}$ is to be valid for
quantum systems with finite range interaction and a spectrum gap,
while for gapless systems other scalings are possible,
$l^{d-1}\log l$ in particular, which is closely related to the
existence of a quantum phase transition in the corresponding
system \cite{Am-Co:08}. This is, however, not simple to prove,
even in the translation invariant case, since the spectrum of many-body quantum
systems is rather complex and is known mostly for 1d exactly
solvable models. On the other hand, there is a simpler model having the
both types of spectrum and the both scalings. These are the
quasi-free fermions described by the Hamiltonian quadratic in the
creation and annihilation operators.
%It arises in condensed matter
%theory and statistical physics (electrons in metals other mean
%field type approximations, exactly solvable spin chains, etc.).
For this Hamiltonian with finite range and translation invariant
coefficients the large-$l$ scaling of the entanglement entropy for
any $d\geq 1$ was established first via the upper and lower bounds
and certain conjectures on the subleading term in the Szeg\"o
theorem for Toeplitz determinants \cite{Gi-Kl:06} and
then rigorously %, by using a rather sophisticated techniques of
%modern operator theory
\cite{So-Co:13}.

All the above concerns the translation invariant systems.
Following a standard paradigm of condensed matter theory, it is
natural to consider a disordered version of the model replacing
the translation invariant coefficients of the fermionic
Hamiltonian by random coefficients, which are translation
invariant in the mean and have fast decaying spatial correlations
\cite{LGP:88}.

The analysis of quadratic fermionic forms reduces to that of a certain one
body Hamiltonian. %determined by the form coefficients.
Thus, in the case of random coefficients we obtain a problem of the theory
of one body disordered systems, which, %, the Anderson
%localization in particular.
%It is easy to show that for the ground state of the fermion
%quadratic form the reduced density matrix corresponding to a domain $\Lambda
%$ of the $d$-dimensional cubic lattice $\mathbb{Z}^{d}$ can be expressed via
%the spectral projection $\theta (\mu - H)$ of $H$, where $\mu$ is the Fermi energy
%and $\theta$ is the Heaviside step function.
however, proves to be quite non-trivial in general. In this
situation and to demonstrate the role of disorder in the
asymptotic behavior of the entanglement entropy without involving
too much technicalities it is natural to use a simple but
non-trivial setting and to ask simpler questions, e.g., on upper
and lower bounds for the disorder averaged entanglement entropy
implying its scaling (see \cite{Gi-Kl:06} for analogous approach
in the translation invariant case), the same for the entanglement
entropy of typical realizations of disorder and/or on the
selfaveraging property of the entropy. Recall that a number of
important characteristics of disordered system (free energy,
magnetization, density of states, conductivity, etc.) possess this
property, i.e., become nonrandom in the macroscopic limit
\cite{LGP:88}. This allows one to deal only with the disorder
averaged characteristics, but not with the whole their probability
distributions. %since the distribution shrinks to
%the delta-function centered at the average.
%Note that the disorder
%averaged entanglement entropy of several spin chains has been
%already studied \cite{Re-Mo:09}). It was found that the entropy is
%also logarithmically divergent at critical points, although with a
%different than in the translation invariant case coefficient in
%front of $\log \;l$.

We will show in this paper that for the free fermions in the
random external field: \ (i) for any $d\geq 1$ the averaged entanglement
entropy possesses the area law scaling $l^{d-1}$; (ii) for
$d=1$ the same in true for all typical realizations of disorder; (iii)
the entropy is not selfaveraging for $d=1$.

\emph{Model}. We consider the system of $N$ lattice spinless fermions with
the parity conserving Hamiltonian
\begin{equation}
\mathcal{H}=\sum_{j,k=1}^{N}A_{jk}c_{j}^{+}c_{k},  \label{HF}
\end{equation}%
where $c_{j}^{+},c_{j},\;j=1,\dots ,N$ are the Fermi operators and $A=\{A_{jk}\}
$ is a hermitian $N\times N$ matrix.
%and $B=\{B_{jk}\}$ is $N\times N$ antisymmetric

%The case, where the indices $j,k$ in (\ref{HF}) are the coordinates of the
%sites of a cube $\Omega ,\;|\Omega |=N$ of the $d$-dimensional cubic lattice
%$\mathbb{Z}^{d}$ arises in various approximations of statistical mechanics and condensed matter
%theory (mean field approximations, BCS theory of superconductivity, etc).%
%To avoid technicalities we consider the "tight-binding" case of (%
%\ref{HF}) where $B=0$ although many results below are valid for  $%
%B\neq 0$ as well (and will be published elsewhere).
By using the Bogolyubov transformation it is easy to find that if $%
K=\{<c_{j}^{+}c_{k}>_{G}\}_{j,k=1}^{N}$ where $<...>_{G}$ is the "Gibbs"
averaging with the "density" matrix
\begin{equation}
\rho =e^{-\mathcal{H}}/Z,\;Z=\mathrm{Tr\ }e^{-\mathcal{H}}/Z,  \label{rf}
\end{equation}%
then
\begin{eqnarray}
K&=&(1+e^{A})^{-1},\;A=-\log K(1-K)^{-1},  \label{ka}\\
S&=&-\mathrm{Tr\ }\rho \;\mathrm{\log }_{2}\rho
=\mathrm{tr\;}h(K),
\label{S} \\
h(x) &=&-x\log _{2}x-(1-x)\log _{2}(1-x),  \label{hx}
\end{eqnarray}%
where $\mathrm{Tr}$ and $\mathrm{tr}$ denote the trace in the $2^{N}$
-dimensional space of $N$ fermions and in the $N$-dimensional one body
configuration space respectively. %and
%\begin{equation}
%h(x)=-x\log _{2}x-(1-x)\log _{2}(1-x), \; 0 \le x \le 1
%\label{hx}
%\end{equation}
%is the binary Shannon entropy.

We choose $A=(H-\mu )/T$ where
%\begin{equation}\label{and}
$H=H_0+V$
%\end{equation}
is the Anderson model, in which $H_0= a \
\Delta$, $a$ is the hopping parameter, $\Delta $ is the discrete
Laplacian, $V=\{V_{j}\}_{j\in \Omega }$ is the random potential,
$\mu $ is the Fermi energy and $T$ is the temperature. Then
(\ref{ka}) implies
\begin{equation}
P=\{P_{jk}\}_{j,k \in \Omega}=K\left\vert _{T=0}\right. =\theta (\mu -H),  \label{K0}
\end{equation}%
where $\theta $ is the Heaviside function. Thus, $P$ is the
orthogonal projection on the ground state of the whole system, the
Slater determinant on the first $n$ eigenstates of $H$, where
$n/|\Omega |=N(\mu )$ and $N(\mu ) $ is the integrated density of
states of $H$. Hence the entropy (\ref{S}) of the whole system is
zero.

Consider now a subsystem of fermions in a sub-cube $\Lambda $ of
$\Omega $, the latter can be the whole $\mathbb{Z}^{d}$. We assume
that $\Lambda$ is centered at the origin and of side length
$l=2m+1 $. Note that the setting is not unambiguous for
indistinguishable
particles and we use its natural version known as the entanglement of modes \cite%
{Pe-In:09}. Then  the corresponding reduced density matrix is $%
\rho_{\Lambda} =e^{-\mathcal{H}_{\Lambda }}/Z_{\Lambda }$, where $\mathcal{H}%
_{\Lambda }$ is the entanglement Hamiltonian \cite{Am-Co:08} given
by (\ref{HF}) with $A=-\log P_{\Lambda }(1-P_{\Lambda })^{-1}$ and
(see (\ref{K0}) and (\ref{S})))
\begin{equation}
S_{\Lambda }=-\mathrm{Tr\ }\rho _{\Lambda }\mathrm{\log }_{2}\rho
_{\Lambda }=\mathrm{tr\;}h(P_{\Lambda }), \; P_{\Lambda
}=\{P_{jk}\}_{j,k \in \Lambda}.  \label{sl}
\end{equation}

\emph{The area law scaling for the disorder averaged entanglement entropy}.
We will show now that if the spectrum of $H$ below $\mu $ is localized, then
the disorder average $\langle S_{\Lambda}\rangle $ scales as $l^{d-1}$ for $%
l \gg 1$. To this end we present an upper and a lower bounds for
$\langle S_{\Lambda }\rangle$, which are asymptotically
proportional to $l^{d-1}$.

We start from bounds for $h$ of (\ref{S}) \cite{Am-Co:08}:
\begin{equation}
\varphi (x)\leq h(x)\leq \sqrt{\varphi (x)},\;\varphi (x)=4x(1-x).
\label{tbou}
\end{equation}
%Denote $\{P_{jk}\}_{j,k\in \mathbb{Z}^{d}}$ the entries of $K^{0}$ of (\ref%
%{K0}). Then $K_{\Lambda }=\{P_{jk}\}_{j,k\in \Lambda }$ and (\ref{tbou})
The bounds and (\ref{sl}) imply%
\begin{eqnarray}
&&L_{\Lambda }\leq S_{\Lambda }\leq U_{\Lambda },\;L_{\Lambda }=4\;\mathrm{tr%
}\Gamma _{\Lambda }\   \label{lub} \\
&&U_{\Lambda }=2\;\mathrm{tr}\sqrt{\Gamma _{\Lambda }},\;\Gamma _{\Lambda
}=K_{\Lambda }(\mathbf{1}_{\Lambda }-K_{\Lambda }).  \notag
\end{eqnarray}%
Use the equality $\sum_{k\in \mathbb{Z}^{d}}|P_{jk}|^{2}=P_{jj}$ valid for
any orthogonal projection to write
\begin{equation}
L_{\Lambda }=4\sum_{j\in \Lambda }\sum_{k\in \overline{\Lambda }%
}|P_{jk}|^{2},  \label{ll}
\end{equation}%
where $\overline{\Lambda }$ the exterior of $\Lambda $.

Note that in the 1d translation invariant case
$P_{jk}=\sin \kappa (j-k)/\pi (j-k)%\end{equation*}%
$ where $\kappa $ is the Fermi momentum and (\ref{expll}) yields
%\begin{equation*}
$L_{\Lambda }\simeq 4\pi ^{-2}\log l,\;l\gg 1 $.
%\end{equation*}%
This is a simple example of the $\log$-scaling in the translation
invariant case. A more involved argument leads to the lower bound
$\sim l^{d-1}\log l $ for any $d\geq 1$ \cite{Gi-Kl:06}
and to the corresponding asymptotic formula \cite{So-Co:13}.

Assume that the potential is independent and identically
distributed (i.i.d.) in different points. Then $\mathbf{\langle
}|P_{jk}|^{2}\rangle =\Pi _{j-k}$, where $\Pi _{j}=\Pi _{-j}$ and
is symmetric in the coordinates $(j_{1},...,j_{d})$, and
(\ref{ll}) implies
\begin{equation}
\langle L_{\Lambda }\rangle =4\sum_{j\in \Lambda }\sum_{k\in \overline{%
\Lambda }}\Pi _{j-k}.  \label{ell}
\end{equation}%
It is easy to find that $\Pi _{j-k}$ is the integral over $\Delta
\times \Delta ,\ \Delta =(-\infty ,\mu ),$ of the current-current
correlator $\langle (\delta (H-E_{1}))_{jk}(\delta
(H-E_{2}))_{jk}\rangle $ determining the a.c. conductivity of free
disordered fermions \cite{Ki-Co:03}.

We will use now a basic rigorous result of the localization of
states of the $d$-dimensional Anderson model, according to which
if the probability distribution of i.i.d. random potential is
smooth enough and either $\mu$ is close enough to the bottom of
the spectrum or the hopping parameter is small enough, then
$\mathbf{\langle }|P_{jk}|\rangle \leq Ce^{-\gamma
|j-k|}$ for some $%
C<\infty $ and $\gamma >0$   (see, e.g. \cite{St:11}). This and
the inequality
$|P_{jk}|\leq 1$ valid for any orthogonal projection imply%
\begin{equation}
\Pi _{j}\leq Ce^{-\gamma |j|}.  \label{loc}
\end{equation}%
The sum over $k$ in (\ref{ell}) consists of $2^{d}-1$ sums such
that
$\binom{d}{\delta },\;\delta =1,...,d$ of them have the coordinates $%
k_{a_{1}},...,k_{\alpha _{\delta }}$ outside the interval $[-m,m]$
and the rest inside the interval. Since the summands of
(\ref{ell}) are non-negative, $\langle L_{\Lambda }\rangle $ is
bounded below by the sums with $\delta =1$ (in fact, the leading
term of $\langle L_{\Lambda }\rangle $ for $l\gg 1$) and then
(\ref{loc}) yields that up to exponential small in $l$ terms
\begin{eqnarray}
&&\langle L_{\Lambda }\rangle \geq 4d\sum_{j\in \Lambda }\sum_{|k_{1}|>m}\Pi
_{|j_{1}-k_{1}|}^{(1)}\simeq c_{-}l^{d-1},  \label{expll} \\
&&c_{-}=8d\sum_{t\geq 1}t\Pi _{t}^{(1)},\;\Pi
_{t}^{(1)}=\sum_{j_{2},...,j_{d}\in \mathbb{Z}^{d-1}}\Pi
_{t,j_{2},...,j_{d}}.,  \notag
\end{eqnarray}%
%
%where $\Pi _{t}^{(1)}$is the sum of $\Pi _{j},\;j=(t,j_{2},...,j_{d})$ over
%all integers $(j_{2},...,j_{d})$.
We write here and below $a_l\simeq b_l$ if $b_l$ is the leading
term of $a_l$ for $l \gg l$.

For the upper bound $U_{\Lambda }$ of (\ref{lub}) we will use the
inequality $\mathrm{Tr}f(M)\leq \sum_{j=1}^{n}f(M_{jj})$ valid for
any $n\times n$ hermitian $M$ and a concave $f$. The inequality is
a version of the Peierls variation principle \cite{Ru:77} with the
only difference is that it is usually formulated for convex $f$,
$e^{-x}$ in particular, thus with the opposite inequality.

Use the inequality with $M=\Gamma _{\Lambda }$ of (\ref{lub}) and $f(x)=%
\sqrt{x}$ to obtain (cf. (\ref{ll}))
\begin{equation}
U_{\Lambda }\leq 2\sum_{j\in \Lambda }\Big(\sum_{k\in \overline{\Lambda }%
}|P_{jk}|^{2}\Big)^{1/2}  \label{ul}
\end{equation}%
and then the Schwarz inequality $\langle \xi ^{1/2}\rangle \leq \langle \xi
\rangle ^{1/2}$ and (\ref{loc}) (cf. (\ref{ell}))%
\begin{equation}
\langle U_{\Lambda }\rangle \leq \sum_{j\in \Lambda }\Big(\sum_{k\in
\overline{\Lambda }}\Pi _{k-j}\Big)^{1/2}<\infty .  \label{eul}
\end{equation}%
Since $\Pi _{jk}\geq 0$, the sum over $k=(k_{1},k_{2,}...,k_{d})\in
\overline{\Lambda }$ is not less than $(2^{d}-1)$ times the sum over $%
|k_{1}|>l$ and $(k_{2},...,k_{d})\in \mathbb{Z}^{d-1}$. This and the
elementary inequality $\sqrt{a+b}\leq \sqrt{a}+\sqrt{b}$ yield up to
exponential small in $l$ terms (cf. (\ref{expll}))%
\begin{equation}
\langle U_{\Lambda }\rangle \lesssim
c_{+}l^{d-1},\;c_{+}=4(2^{d}-1)\sum_{j=0}^{\infty }\Big(\sum_{k=1}^{\infty
}\Pi _{k+j}^{(1)}\Big)^{1/2}.  \label{expul}
\end{equation}%
Note that $c_{\pm }$ of (\ref{expll}) and (\ref{expul}) are finite
view of (\ref{loc}). %, i.e., the complete  localization of spectrum.
This and (\ref{lub}) prove the validity of the area law scaling
$\langle S_{\Lambda }\rangle \sim l^{d-1}$ for the averaged
entanglement entropy of free disordered fermions.
%(the validity would mean that $\langle S_{\Lambda %}\rangle $ is
%asymptotically equals $b(2l+1)^{d-1}$).
For similar results on disordered oscillators see \cite{Na-Co:13}.
%\begin{figure}
%\epsfig{width=8cm,file=fig2.eps}
%\caption{The histograms of distributions of the lower (triangles)
%and upper (squares) bounds (\ref{lb}) and (\ref{ub}).
%The hopping parameter is 1/10 the potential is uniformly distributed over [-1,1] and $\mu=-0.25$
%The system ($\Omega$) size is 30000 and the subsystem ($\Lambda$) size is 1500.
%System size $L$=3000, $\Delta$=[-0.25,0.25], $\alpha=0.1$. $L/l=20$
%}
%\label{fig1}
%\end{figure}

\emph{Bounds for the 1d entanglement entropy on typical realizations of
disorder}. We will write $L_{l}$ and $U_{l}$ for $L_{\Lambda }$ and $%
U_{\Lambda }$ and $\Lambda =[-m,m]$ and $l=2m+1$. We have
\begin{eqnarray}
&&L_{l}=4\sum_{|j|\leq m}\sum_{|k|>m}|P_{jk}|^{2}=L_{l}^{+}+L_{l}^{-},
\label{prid} \\
&&L_{l}^{+}=4\sum_{|j|\leq
m}\sum_{k>m}|P_{jk}|^{2},\;\;L_{l}^{-}=4\sum_{|j|\leq
m}\sum_{k<-m}|P_{jk}|^{2},  \notag
\end{eqnarray}%
thus,
\begin{eqnarray}
&&L_{l}^{\pm }=\mathcal{L}_{l}^{\pm }+R_{l}^{\pm },\;\mathcal{L}_{l}^{\pm
}=4\sum_{j=\pm m}^{\mp \infty }\sum_{k=\pm (m+1)}^{\pm \infty }|P_{jk}|^{2},
\notag \\
&&R_{l}^{\pm }=4\sum_{j=\mp (m+1)}^{\mp \infty }\sum_{k=\pm (m+1)}^{\pm
\infty }|P_{jk}|^{2},  \label{tlp}
\end{eqnarray}%
According to (\ref{loc}), $\mathbf{\langle }R_{l}^{\pm }\rangle
\leq C_{1}e^{-\gamma _{1}l}$ where $C_{1}<\infty $ and $\gamma
_{1}>0$. This,
the Tchebyshev inequality and the Borel-Cantelli lemma \cite{Fe} imply that $%
R_{l}^{+}$ vanishes with probability 1 as $l\rightarrow \infty $, i.e.,
%\begin{equation}
$L_{l}^{+}\simeq \mathcal{L}_{l}^{+},\;l\gg 1$ with probability 1.
%\label{lill}
%\end{equation}%
Introduce the shift operator $T$: $(TV)_{j}=V_{j+1}$. Writing the
Anderson Hamiltonian as $H(V)$ to make explicit its dependence on
$V$, we find that $H_{jk}(T^{a}V)=H_{j+a,k+a}(V)$. This and
(\ref{K0}) imply the same for $\{P_{jk}\}_{j,k\in \mathbb{Z}}$,
thus \ $\mathcal{L}_{m}^{+}(V)=%
%4\sum_{j=-\infty }^{0}\sum_{k=1}^{\infty}|P_{j+l,k+l}(V)|^{2}=
\mathcal{L}_{0}^{+}(T^{l}V).$ %\end{equation*}%
%where%
%\begin{equation}
%\mathcal{L}^{+}(V)=4\sum_{j=-\infty }^{0}\sum_{k=1}^{\infty }|P_{jk}(V)|^{2}.
%\label{tplus}
%\end{equation}%
The terms of the series in (\ref{tlp}) for $\mathcal{L}_{0}^{+}$ are
nonnegative random functions, thus the series is convergent with probability
1 if the series of its averages is convergent. %\begin{equation*}
%\sum_{j=0}^{\infty }\sum_{k=1}^{\infty }\langle %|P_{-j,k}|^{2}\rangle <\infty .
%\end{equation*}%
This is again guarantied by (\ref{loc}). Thus, $\mathcal{L}_{0}^{+}$ of (\ref%
{tlp}) is well defined and we have $L_{l}^{+}\simeq
\mathcal{L}_{0}^{+}(T^{m}V),\;l\gg 1$ with probability 1. Likewise
$L_{l}^{-}\simeq \mathcal{L}_{0}^{-}(T^{-m}V),\;l\gg 1$ and
\begin{equation}
L_{l}\simeq \mathcal{L}_{0}^{+}(T^{m}V)+\mathcal{L}_{0}^{-}(T^{-m}V),\;l\gg
1.  \label{lb}
\end{equation}%
with the same probability.

A similar argument yields (cf. (\ref{lb}))
\begin{eqnarray}
&&U_{l}\lesssim \mathcal{U}_{0}^{+}(T^{m}V)+\mathcal{U}_{0}^{-}(T^{-m}V),\;l%
\gg 1,  \notag \\
&&\mathcal{U}_{0}^{\pm }(V)=2^{3/2}\sum_{j=0}^{\mp \infty }\Big(\sum_{k=\pm
1}^{\pm \infty }|P_{k,j}|^{2}\Big)^{1/2}.  \label{ub}
\end{eqnarray}%
\begin{figure}[tbp]
\epsfig{width=8cm,file=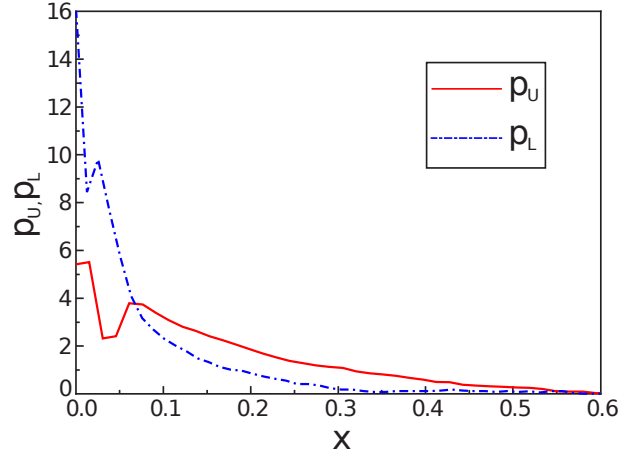} \caption{The probability
distributions $p_U(x)$ and $p_L(x)$ of the lower
(\protect\ref{lb}) and upper (\protect\ref{ub}) bounds obtained
from numerical data on 15000 realizations of disorder. The hopping
parameter $a$ of the Anderson model is $1/10$, the potential is
uniformly distributed over [-1,1], $ \protect\mu=-0.25$, the
system size $L=30000$ and the subsystem size $l=1500$.}
\label{fig1}
\end{figure}
Figure~\ref{fig1} presents our numerical results on the
probability distributions $p_{L}(x)$ and $p_{U}(x)$ of the lower (\ref{lb})
and upper (\ref{ub}) bounds, the latter is with the
optimal exponent $\log _{2}$ instead of $1/2$ (see Remark (i)
below).
%They are obtained from the numerical data on 10000 realization of disorder.
%for the hopping parameter 1/10 and the potential uniformly distributed
%over [-1,1].The system ($\Omega$) size is 30000 and the subsystem ($\Lambda$) size is 1500.
It is important that $p_{L}$ and $p_{U}$ are non-zero on the
practically the same intervals. This implies that the entanglement
entropy $S_{l}$ depends nontrivially on the realizations of
disorder even if $l\gg 1$, i.e., $S_{l}$ \emph{is not
selfaveraging}. Indeed, if it were selfaveraging, i.e.,
$S_{l}\simeq S,\ l\gg 1$ for a non-random $S$, then the whole
interval where the probability density $p_U$ of the upper bound
(\ref{ub}) is non-zero would lie on the right of $S$, while the
whole interval where the probability density $p_L$ of the lower
bound (\ref{lb}) is not zero would lie on the left of $S$. Thus,
these two probability densities would not overlap.

Besides, it follows from the analysis of numerically obtained
probability distributions of $U_l$ and $L_l$ with growing $l$ that
they become independent of $l$ (saturate) for $l\gg 1$. This can
be explained as follows. Since the random potential is independent
in different points, the first two terms of the r.h.s. of
(\ref{lb}) and (\ref{ub}) has to be also statistically independent
for $l\gg 1$, and since the potential is translation and
reflection symmetric in the mean, the probability distributions of
these terms are independent of $m$ and identical. Hence, for $l\gg
1$ the probability distributions $p_L$ and $p_U$ of (\ref{lb}) and
(\ref{ub}) are the convolutions of $l$-independent probability
distributions of $\mathcal{L}_{0}^{\pm }$ and
$\mathcal{U}_{0}^{\pm } $ and this was also checked numerically.

It worth mentioning that our numerical results do not allow us to
conclude that the probability distribution of the entanglement
entropy $S_l, \; l \gg 1$ is concentrated on a finite interval,
hence that the random function $S_l$ is bounded by a non-random
constant for $l \gg 1$ on the typical realizations of disorder. In
fact, this seems unlikely, hence one has to expect that for every
typical realization the random function $S_l$ assumes arbitrary
large values $S_{l_{n}}$ on an infinite and depending on
realization sequence of values $l_n$ of $l$. However, these would
be just rather rare peaks of randomly fluctuating entanglement
entropy (\ref{sl}) but not its "regular" asymptotics.

\emph{Remarks}. (i) The bound $\sqrt{\varphi}$ in (\ref{tbou}) can
be replaced by a tighter one $\varphi^\alpha, \; \alpha=\log 2$.
(ii) Analogous results are valid for the R\'enyi entropy
$R_{\alpha}=(1-\alpha)^{-1} \mathrm{Tr} \log_2
\rho_{\Lambda}^{\alpha}$
%=(1-\alpha)^{-1} %\mathrm{tr}\log_2(K_\Lambda^\alpha + (1-K_\Lambda)^\alpha),$$
%$\alpha>0$,
reducing to the von Neumann entropy (\ref{sl}) for $\alpha=1$.
(iii) The above results are based on (\ref{loc}) manifesting the
localization for the corresponding one-body problem. It follows
from \cite{Ji-Kr:13} that an analogous bound holds for 1d
Schrodinger operator with certain incommensurate potentials. Thus,
the entanglement entropy of 1d free fermions in the corresponding
external fields is also bounded. (iv) We have discussed the area
law for the Fermi energy lying in the localized spectrum of the
Anderson model. The case, where the Fermi energy is in a gap is
much simpler. Here an analog of (\ref{loc}) can be obtained by
writing (\ref{K0}) as the contour integral of the exponentially
decaying Green's function. (v) One can ask on the asymptotics of
the entanglement entropy for non-zero temperature (thermal
entanglement). In this case the leading term of the entropy is
proportional to $l^d$ with a non-random coefficient and there are
certain random or incommensurate subleading terms of various
scaling (a stochastic analog of Szeg\"o theorem \cite{Pa:14}).

\emph{Conclusion}. We have shown that for the free fermions in the
random external field the averaged entanglement entropy of the
$d\geq 1$ dimension cube of side length $l$ is bounded from above
and from below by $c_{\pm }l^{d-1}$. The result suggests the
validity of the area law "in the mean" even in the gapless case
for disordered free fermions. This has to be compared with the
results for the translation invariant case, where the entropy
scales as $l^{d-1}\log l$ \ \cite{Am-Co:08}, and with those of a
series of works (see \cite{Re-Mo:09} for a review) in which, by
using a strong disorder version of the real space renormalization
group, it was found that the averaged entropy at critical points
of certain disordered spin chains scales as in the non-random
case, although with a different prefactor of $\log l$. This could
be an indication of the difference of the origin of the area law
for disordered spin chains and disordered free fermions where
there is no interaction and a non-trivial entanglement is due to a
pure "kinematic" effect of Fermi statistics, hence simple formulas
(\ref{ka}) -- (\ref{hx}).
%which have more complex structure of
%multipoint correlation functions.

We have also obtained bounds for the $d=1$ entanglement entropy of
all typical realizations of disorder. The bounds do not imply in
general that the entropy of typical realizations is bounded for
$l\gg 1$ "uniformly" in realizations, i.e., by a non-random
constant. However, we show numerically that the bounds have a non-trivial $l$%
-independent for $l\gg 1$ and overlapping probability
distributions (see Fig.~\ref{fig1}) manifesting that the
entanglement entropy is not selfaveraging, i.e., has non vanishing
random fluctuations for $l\gg 1$.

Our results can be viewed as an indication of an important role of
disorder in the entanglement in extended systems, similarly to its
role in condensed matter (Anderson localization) and phase
transitions (rounding effects). This seems to be especially
interesting in the dimension 1, where the Anderson localization is
the case for arbitrary small disorder and all energies
\cite{LGP:88}. The results can also be used in the elaboration of
the DMRG method (see \cite{Am-Co:08,Pe-In:09} for reviews) for
disordered systems (finite-size effects, possible regularization
tools, etc.)

\end{document}